\documentclass{article}
\usepackage{graphicx,graphics,amsmath,natbib}

\begin{document}

\title{Rms-flux relation in the optical fast variability data of BL Lacertae object S5 0716+714}
\author{Gabriela Raluca Mocanu, Bulcs\'{u} S\'{a}ndor\\
\small{Faculty of Physics, Babes-Bolyai University, No. 1
Kog\u{a}lniceanu Street,}\\ \small{400084, Cluj-Napoca, Romania,
gabriela.mocanu@ubbcluj.ro}}

\date{}

\maketitle

\begin{abstract}
The possibility that BL Lac S5 0716+714 exhibits a linear root mean square (rms)-flux
relation in its IntraDay Variability (IDV) is analysed. The results may
be used as an argument in the existing debate regarding the source
of optical IDV in Active Galactic Nuclei. $63$ time series in
different optical bands were used. A linear rms-flux relation at a
confidence level higher than $65\%$ was recovered for less than
$8\%$ of the cases. We were able to check if the magnitude is log-normally distributed for eight timeseries and found, with a confidence $> 95 \%$, that this is not the case.\\ \textbf{keywords}: BL Lacertae objects: individual(S5 0716+714) --
X-Rays: rms-flux relation -- optical: IDV.
\end{abstract}

\section{Introduction}

The linear rms-flux relation for variability in X-Rays has been
established observationally to be almost ubiquitous in a wide
range of objects, starting from X-Ray Binaries (XRBs) to
super massive black holes in the active regime (Active Galactic
Nuclei)~\cite{art:5u,art:6u,art:45s}. One of its major importance
lies in the fact that this relation is observed even while the Power Spectral Distribution (PSD)
shape of the light curve is stationary, implying that this
relation is a fundamental property of the variability
process~\cite{art:5u}. The physics behind this relation is that
the sources become more variable as they get brighter and it has
been shown that it predicts a log-normal distribution of fluxes,
implying that the variability process is non-linear~\cite{art:5u}.~\footnote{Interestingly, the same linear rms-flux relation is
exhibited by solar flares, even though the Sun does not have an
accretion disk~\cite{art:5z}. However, the flux distribution of
solar flares is well fitted by a power law, a feature explained by
(additive) SOC (see e.g.~\cite{art:15m}), while usually (multiplicative) processes which
give a linear rms-flux also give a log-normal flux distribution.}

The optical/UV and X-Ray continua are thought to originate in two
physically distinct but nevertheless related
regions~\cite{art:2y}, the optical-X-Ray emissions being
connected through reprocessing in the
disk~\cite{art:41k,art:12s}, especially in a two phase thermal
model~\cite{art:43s}. This framework is based on observational
data and theoretical arguments, e.g., the ratio of rest frame UV to
X-Ray emission increases with accretion
rate~\cite{art:35k,art:2y}. This framework can be relaxed and a
better fit of the data can be obtained with multicomponent models,
based on the working assumption that only a fraction of the
optical variability is a results of reprocessing in the disk,
while the other fraction is consistent with emission from a
jet\footnote{It is interesting to note that if the response of the
optical reprocessing region to incoming X-Ray flux is non-linear,
i.e. $f_{opt} \sim f _{X}^4$, a large portion of the statistical
properties of rapid (less than 100 s for object XTE J1118+480)
variability can be explained~\cite{art:35h}.} (e.g.
\cite{art:35h} or the energy reservoir model of \cite{art:46m}).

Short time-scale variability in X-Ray may be explained by the
promising disk-model of \cite{art:29l}, which also naturally explains
both the PSD of the light curve and the linear rms-flux
relation~\cite{art:16a,art:45s}. In fact, existence of a linear rms-flux relation is considered as an additional evidence for the fluctuations accretion scenarios because it suggests that the variability originates in the accretion flow~\cite{art:17a}.

Almost ubiquitous in BL Lac objects~\cite{art:19w} and present in
other AGN flavours, IntraDay Variability (IDV, variability on timescales less that one day) in the optical
regime is still an issue of debate \footnote{For variability on longer timescales, of the order of the thermal timescale, the source of optical/UV quasar variability may be consistently placed in the accretion disk. This conclusion is based on reverberation mapping arguments and on the strong correlation between both the characteristic timescale and rms of variability with the black hole mass (see e.g.~\cite{art:49k} and references therein).}. One of the debated aspects is the location of the source, where one has to take into account that  microvariability is known to be stronger in radio loud quasars~\cite{art:49k}. This enhanced level suggests that microvariability is due to processes in a jet. It was shown, for a sample of  117 radio  quiet sources, that the probability of microvariability is a smoothly varying function of the radio loudness~\cite{art:32c}. Arguments of this type have led to simulation efforts to asses detection probabilities. Results for 10 radio quiet quasars, with confirmed intranight variability and with available X Ray data to constrain some of the features of the model, favour the assumption that the variability was caused by the presence of a weak jet component~\cite{art:31c}. However, recent simulation results of~\cite{art:49k} suggest that if one interprets the optical flux fluctuations as resulting from thermal fluctuations that are driven by an underlying disk-based stochastic process, one can obtain microvariability consistent with observations of radio quiet quasars. 

Although there are isolated objects for which close correlation
between the UV and X-Ray short time variability has been reported,
e.g. for PKS 2155-304~\cite{art:5b,art:1u}, or for
3C390.3~\cite{art:12g}, and even some low mass objects for which
a linear rms-flux relation exists in the optical (accreting white
dwarfs show linear rms-flux in optical variability on a 10 day
time scale~\cite{art:45s} and~\cite{art:26g} found optical linear
rms flux for 3 XRBs) there is no consistent correlation between
variability in optical-UV and X-Rray on short time scales for
AGN~\cite{art:32k}. \cite{art:33l} found a linear rms-flux
relation in the U band variability on scales of tens of days, for
NGC 4151.

We propose that a valuable argument in this debate might be
offered by analysing whether or not optical IDV in AGNs shows a
linear rms-flux relation\footnote{For the timseries discussed in this work, evaluating the rms-magnitude relation serves the same purpose as the rms-flux. This can be seen very easily if one plots  flux vs. magnitude for the typical parameters used in this paper.  This is why throughout the body of the paper we consider that the magnitude-flux transformation is linear and discuss our results as if the \emph{behaviour} of the magnitude is equivalent to that of the flux.}.

To our knowledge, there was no such systematic study so far of
optical data from AGN that have already shown a linear
rms-flux in X-Ray variability. In this work a first attempt is
initiated, specifically for the BL Lac S5 0716+714. Ideally, such
a study should be based on analysing simultaneous optical and
X-Ray data of the object. This is not always possible so in the
present study most of the optical data analysed does not have
simultaneous X-Ray data counterpart.

BL Lac S5 0716+714, historically identified as a radio source~\cite{art:3q}, has received a lot of attention in the
literature and for our purposes we mention the work of
\cite{art:7n}, who study data series in the V, R, I optical bands,
collected during Jan 1999 - April 2001. They analysed whether or
not there is a correlation between the source magnitude and the
amplitude of the IDV, and they found no such correlation.

After a short description of the algorithm used for the analysis,
we present the results for a number of 63 optical IDV data series
for BL Lac S5 0716+714 (Section~\ref{sect:data}). Conclusions and cautionary notes
follow in Section~\ref{sect:conclusions}.

Although we focus on a limited number of datasets probing only IDV, the same algorithm applied to observations of longer time scales of variability might show very different results. There are at least two reasons for this. The first might be simply that IDV and longer time scale variations are of different nature. The second might be that the quality of long term observations is better and a hypothetical linear rms-flux relation in the optical could unveil.

\section{Analysis of data \label{sect:data}}

The root mean square deviation of a measured variable $X$ is
calculated as~\cite{art:5u}

\begin{equation}
\sigma _{rms} = \sqrt{\frac{1}{N-1}\sum_{i=1}^N \left(  x_i -
\overline{x}\right )^2},
\end{equation}
where the segment of observations is comprised of $N$ data points,
$x_i$ is the value of $X$ at a time labelled by $i$ and
$\overline{x}$ is the medium value of $X$ on this observational
segment. Usually X-Ray data points are sampled with a very large
frequency and there is a discussion whether for these data it
would be better to use this method not on the light curve, but on
its PSD~\cite{art:6u}. For the optical data, we will perform this
analysis directly on magnitude measurements.

The function investigated is $\overline{x} (\sigma _{rms})$, where
$\overline{x}$ is the magnitude for the optical data or the
counts/second for the X-Ray data. The number of data points to
mediate over so as to obtain $\overline{x}$ is $10$ (a 10 data
point/bin binning procedure).

\subsection{Simultaneous optical and X-Ray observations}

The cases for which optical and X-Ray data were available for the
same event are investigated first.

Results are presented in Table~\ref{tab:fitsSimultaneous}, noting a reference of at least one article where the time
series has been previously analysed and where important details such as observational setup, error analysis and weather conditions may be found. The Table contains information about the date when the timeseries were recorded (first column), the filter (second column), the slope (third column) and intercept (fourth column) of the best linear fit and the value of the $R^2$ goodness-of-fit statistics of the fit to the data (fifth column). 

\begin{table}[h!]
\caption{Fits for the rms-magnitude relation for optical variability,
in the case when simultaneous X-Ray variability data exists, and a
X-Ray linear rms-flux relation has been found. For details see e.g. \cite{art:4o} for the first set, \cite{art:11f} for the second set, and \cite{art:24g} for the third set of data.} \label{tab:fitsSimultaneous}
\begin{tabular}{ccccc}\hline
   Date & Band &  slope & intercept & $R^2$ \\ \hline
    & & & & \\ \hline
    28-03-1996 \\keV & 0.1-2.4  & 1.169 & 0.047 & 0.776 \\ \hline
    28-03-1996  & $R$ & -0.317 & 1.004 & 0.02 \\ \hline

             &  &  & & \\ \hline
    5-04-2004,\\Fig.~\ref{fig:11fX10} keV & 1.5-3.0  & 1.132 & 0.707 & 0.915 \\\hline
    3-04-2004& $V$ & -2.269 & 1.314 &  0.24 \\ \hline
    5-04-2004 ,\\Fig.~\ref{fig:11fOpt10} & $V$  & 0.667 & 0.941 &  0.491 \\ \hline
    6 April 2004 & $V$ & 4.752 & 0.822 & 0.534 \\ \hline
       &      &   & & \\\hline
     7-12-1994 \\keV & 1.5-3.0  & 1.668 &  0.01 &  0.6685  \\ \hline
     7-12-1994 & $R_c$ & 3.873 & 14 & 0.215 \\ \hline
     7-12-1994 & $F98$ & 4.924 & 13.74 & 0.056 \\ \hline
     7-12-1994 & $V$   & -0.733 & 14.26 & 0.178 \\ \hline
\end{tabular}
\end{table}

\begin{figure}[!h]
\includegraphics[scale=0.7]{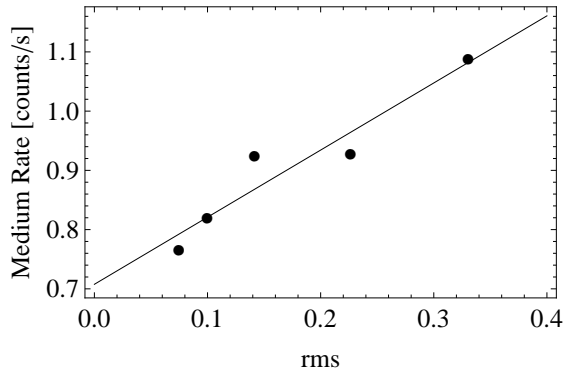}
\caption{For 5 April 2004 the count rate vs. rms spread for X-Rays.}\label{fig:11fX10}
\end{figure}

\begin{figure}[!h]
\includegraphics[scale=0.7]{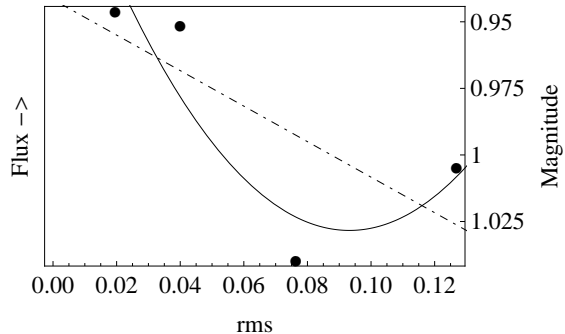}
\caption{For 5 April 2004 the magnitude vs. rms spread for optical (data taken from~\cite{art:11f}). In this case,
the linear fit (dot-dashed) agrees very poorly with the data,
while a second order polynomial fit (full line) produces an $R^2 =
0.8021$.}\label{fig:11fOpt10}
\end{figure}

As an example for the fits, the rms-flux spread for 5 April 2004
both in X-Rays and optical are shown in Figs.~\ref{fig:11fX10} and \ref{fig:11fOpt10}. It is
clear that while for the X-Ray case a linear fit agrees very well
with the points, for the optical case a higher order polynomial is
needed, in this case a second order one. A cautionary note is necessary at this point: there are timeseries amongst those we analysed for which deviation from an "acceptable" linearity was due to only a finite number of points. This is especially obvious in Fig.~\ref{fig:11fX10}, where if one would remove lower point in the plot, a linear fit would work well.

For the date 5 April 2004, where we have almost simultaneous measurement for optical and X-Ray (e.g. Fig. 4 of \cite{art:11f}), we calculate the sample correlation coefficient $r$ between the two timeseries as 

\begin{equation}
r = \frac{\sum _{i=1}^{i=n} \left ( x_i - \bar{x} \right ) \left ( y_i - \bar{y} \right )}{\sqrt{\sum _{i=1}^{i=n}\left ( x_i - \bar{x} \right )^2 \sum _{i=1}^{i=n} \left ( y_i - \bar{y} \right )^2 }},\label{eq:corr}
\end{equation}
where $x$ denotes the optical timeseries, expressed in magnitude and $y$ denotes the X-Ray timeseries, expressed in counts per second\footnote{Since we only care about variations, i.e. the behaviour of relative values, the conversion factors are unimportant as long as the conversion is linear.}. We obtained that $r=0.3638$. A crystal clear diagnosis would have been to have $\lvert r \rvert \in\{0,1\}$, which would have meant anti-correlated (for $-1$), complete uncorrelated (for $0$) and correlated samples (for $1$)~\footnote{Note that because magnitude varies inversely proportional to flux, the numerical value for the correlation coefficient for the optical flux-X-Ray flux timeseries is $r=-0.3638$.}. This value of $r$ does not provide any insight in its own, but in the context of our other data, it also points towards the fact that optical IDV does not show a linear rms-flux relation.

\subsection{Just optical data}

The rms-magnitude relation in optical IDV might prove an insightful
tool for physical analysis of accretion in AGN. With this in mind,
a summary of the best linear fits for a number of IDV data series is
given in Tables~\ref{tab:fits1}, \ref{tab:fits2} and \ref{tab:fits3}. The structure of the Tables is similar to that of Table~\ref{tab:fitsSimultaneous}.

\begin{table}[h!]
\caption{Fits for the rms-magnitude relation for optical variability. Details may be found in \cite{art:7n} for the first group, \cite{art:29m} for the second group, \cite{art:25c} for the third group and \cite{art:11l} for the fourth group of data.}
\label{tab:fits1}
\begin{tabular}{ccccc}\hline
   Date & Band  & slope & intercept  & $R^2$ \\ \hline
  & & & & \\ \hline
    2-01-2000  & $V$  & 9.709 & 13.91 & 0.212 \\ \hline
    12-01-2000 & $V$  & -10.01 & 13.54 & 0.158 \\ \hline
    25-01-2000 & $V$  & -2.152 & 14.03 & 0.016 \\ \hline
    26 Jan 2000 & $V$  & 3.544 & 13.49 & 0.266 \\ \hline
 & & & & \\ \hline
    25-02-2003 & $R$ & -3.549 & 13.07 & 0.121 \\ \hline
 &  & & & \\ \hline
    5-03-2003 & $B$  & -4.213 & 0.398 & 0.484 \\ \hline
    6-03-3003 & $B$  & 0.589 & 0.23 & 0.074 \\ \hline
    7-03-2003 & $B$   & -2.371 &  0.651 & 0.075\\ \hline
    8-03-2003 & $B$   & 1.593 & 0.043  & 0.026 \\ \hline
    9-03-2003 & $B$  & 3.939 &  -0.206  & 0.154 \\ \hline
  &  & & & \\ \hline
    10-01-2007 & $R$& 283.6989 &  -4.8039 & 0.2884 \\ \hline
    12-01-2007 & $R$ & -193.6 & 27.12  & 0.201 \\ \hline
    23-02-2007 &  $R$ & 208.2 &  2.183  & 0.564 \\ \hline
    19-03-2007 & $R$  & 169.3 &  3.354  & 0.207 \\ \hline
    20-03-2007 & $R$  & 13.13 & 13.35  & 0.021 \\ \hline
\end{tabular}
\end{table}

\begin{table}[h!]
\caption{Fits for the rms-magnitude relation for optical variability, data from \cite{art:42m,art:3p}.}
\label{tab:fits2}
\begin{tabular}{ccccc}\hline
   Date & Band  & slope & intercept & $R^2$ \\ \hline
    2454824 & B &  -0.132 & 1.896 & 0.703 \\ \hline
    2454826 & B &  -2.562 & 14.27 & 0.04 \\ \hline
    2454828 & B &   0.01 & - 0,149 & 0.7991 \\ \hline
    2454866-\\2454867 & B   &0.088 &- 1.217 & 0.253 \\ \hline
    2454871-\\2454872 & B   & 0.015 & - 0.199 & 0.068 \\ \hline
    2454872-\\2454873 & B    &-0.023 &  0.359 & 0.045 \\ \hline
    2454765 & V   & -0.005 & 0.081 & 0.006 \\ \hline
    2454766 & V   & -0.014 & 0.201 & 0.055 \\ \hline
    2454767 & V   & -0.046 & 0.633 & 0.287 \\ \hline
    2454770 & V   & 0.025  & -0.348 & 0.153 \\ \hline
    2454824 & V   & -0.129 & 1.787 & 0.766 \\ \hline
    2454825 & V   & 0.014  & -0.194 & 0.061 \\ \hline
    2454826 & V   & -0.003 & 0.05 & 0.005 \\ \hline
    2454828 & V   & 0.118  & -1.657 & 0.173 \\ \hline
    2454829 & V   & 0.042   & -0.579 & 0.297\\ \hline
    2454830 & V   & -0.007 & 0.101 & 0.005 \\ \hline
    2454865-\\2454866 & V    &-0.025 & 0.359 & 0.077 \\ \hline
    2454866-\\2454867 & V    & 0.071 & -0.952 & 0.229 \\ \hline
    2454871-\\2454872 & V    & 0.043 & -0.581 & 0.168 \\ \hline
    2454872-\\2454873 & V    & -0.15 & 0.225 & 0.014 \\ \hline
\end{tabular}
\end{table}

\begin{table}[h!]
\caption{Fits for the rms-magnitude relation for optical variability, data from \cite{art:42m,art:3p}.}
\label{tab:fits3}
\begin{tabular}{ccccc}\hline
   Date & Band  & slope & intercept & $R^2$ \\ \hline
    2454765 & R    & -0.025 & 0.338 & 0.12 \\ \hline
    2454766 & R    & -0.025 & 0.339 & 0.197\\ \hline
    2454767 & R    & -0.143 & 1.895 & 0.82 \\ \hline
    2454770 & R    & 0.026  & 0.341 & 0.154 \\ \hline
    2454824 & R   & -0.121 & 1.623  & 0.555\\ \hline
    2454825 & R   & 0.064 & -0.843 & 0.426 \\ \hline
    2454826 & R   & -0.043 & 0.579 & 0.231 \\ \hline
    2454828 & R    & 0.074 & -1.009 & 0.712 \\ \hline
    2454829 & R    & 0.021 & -0.285 & 0.039 \\ \hline
    2454830 & R    & 0.042 & -0.561 & 0.055 \\ \hline
    2454865-\\2454866 & R   & -6.567 & 1.38 & 0.179 \\ \hline
    2454866-\\2454867 & R   & 0.084 & -1.088 & 0.06 \\ \hline
    2454871-\\2454872 & R   & 0.039 & -0.513 & 0.234 \\ \hline
    2454872-\\2454873 & R   & -0.028 & 0.394 & 0.04 \\ \hline
    2454824 & I  & -0.093 & 0.014 & 0.185 \\ \hline
    2454825 & I  & -0.019 & 0.007 & 0.022 \\ \hline
    2454826 & I  & 0.017  & 0.005 & 0.019 \\ \hline
    2454828 & I  & 0.095  & -0.021 & 0.334 \\ \hline
    2454830 & I  & 0.109  & 0.0 & 0.389 \\ \hline
    2454866-\\2454867 & I   & 0.041 & -0.044 & 0,002 \\ \hline
    2454871-\\2454872 & I   & 0.028 & 0.007 & 0.14 \\ \hline
    \end{tabular}
\end{table}

\begin{figure}[!h]
\includegraphics[scale=0.4]{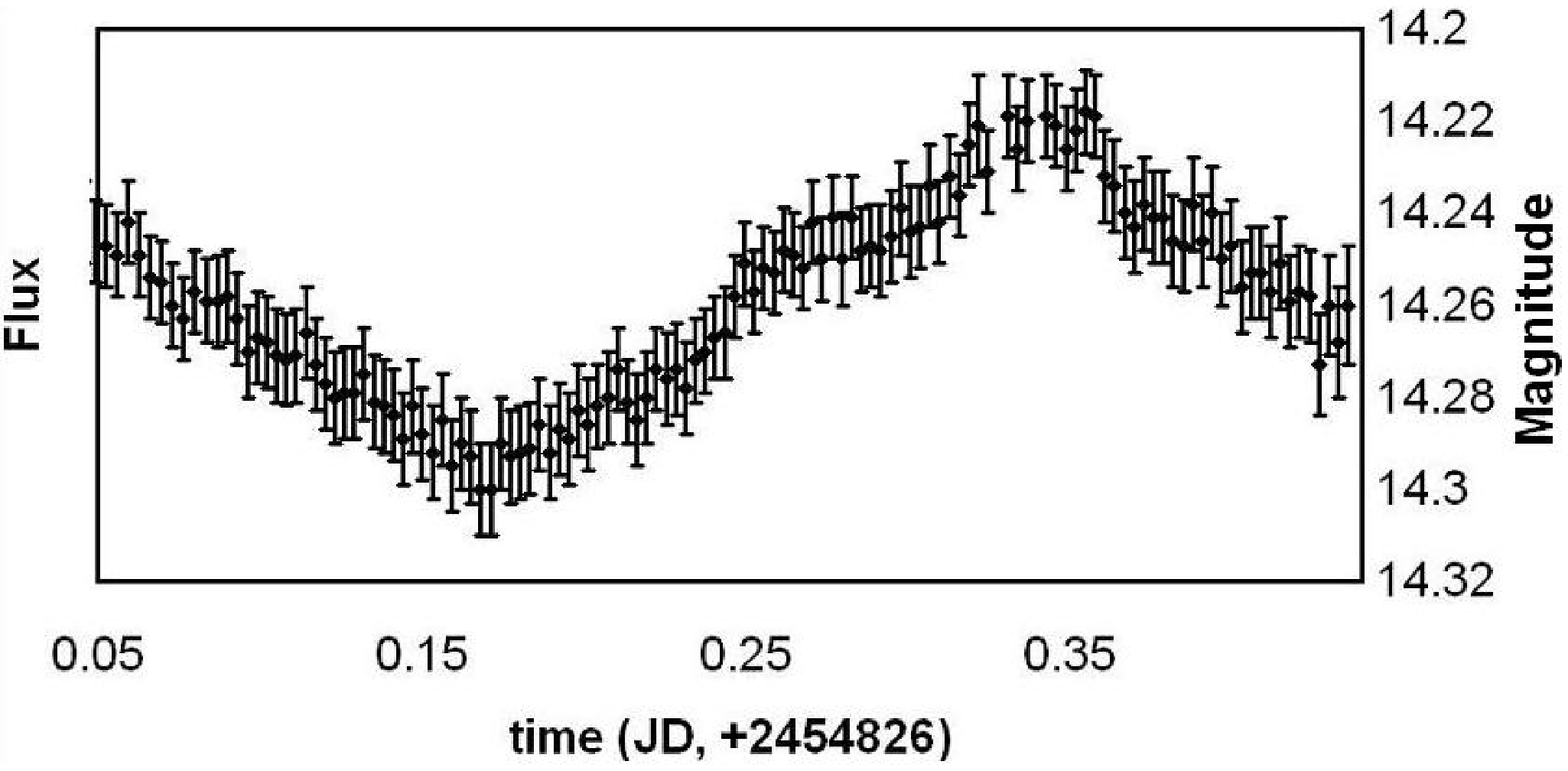}
\caption{the light curve of B 2454826, with error bars plotted for the magnitude measurements. An increment of $0.1$ on the time axis corresponds to $144$ minutes.}\label{fig:B2LC}
\end{figure}

\begin{figure}[!h]
\includegraphics[scale=0.8]{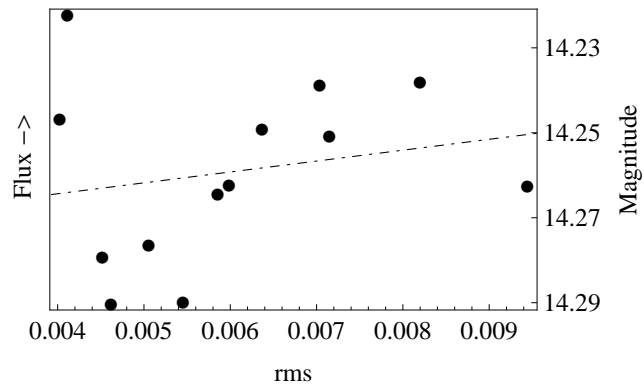}
\caption{The rms-magnitude spread, with a linear fit superimposed for  B 2454826 (data taken
from~\cite{art:42m}).}\label{fig:B210}
\end{figure}

\begin{figure}[!h]
\includegraphics[scale=0.4]{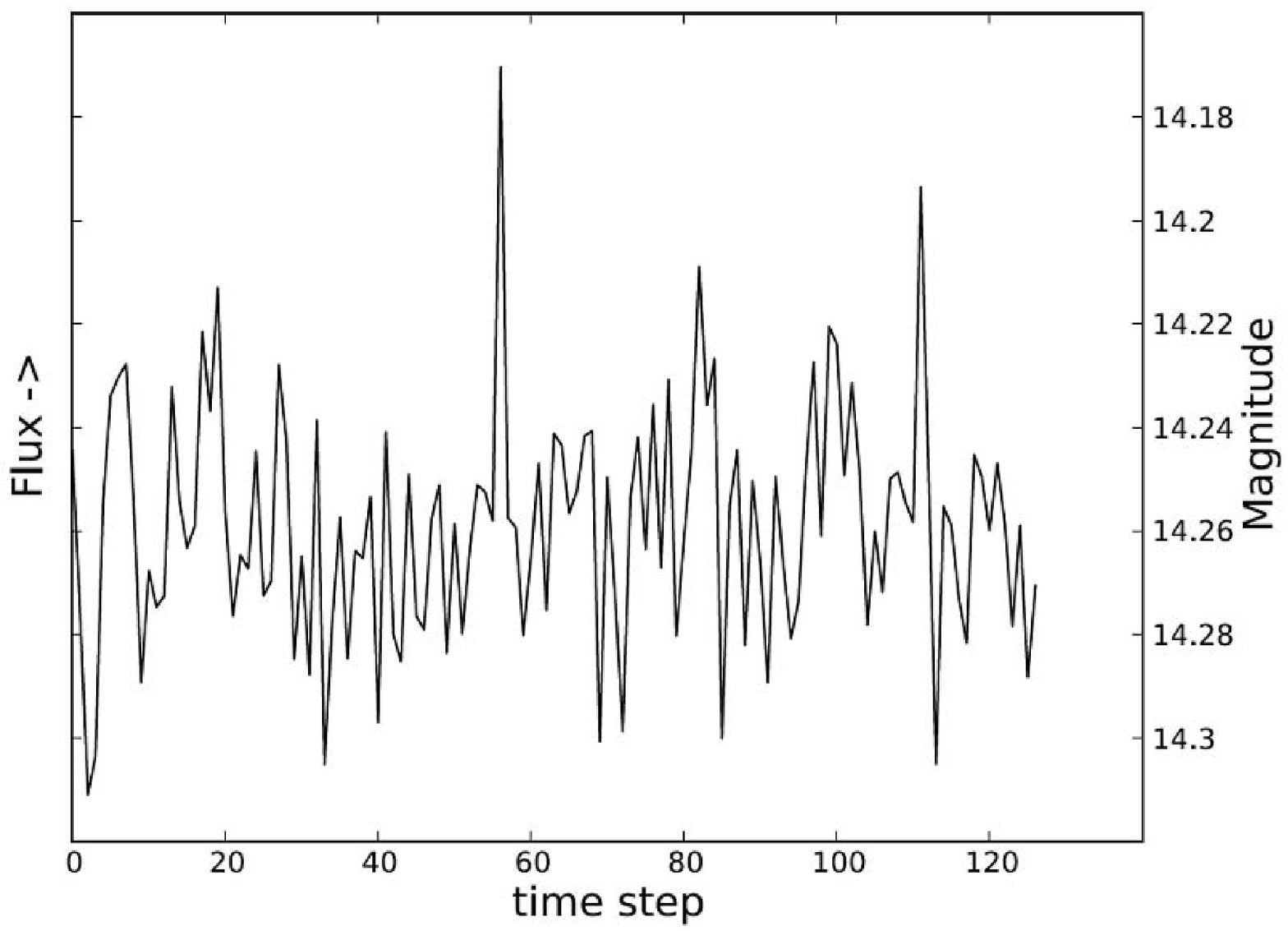}
\caption{Simulated light curve with the same characteristics as the B 2454826 light curve, if the process would be log-normally distributed (see Section~\ref{sec:testLog}).}\label{fig:B2sim}
\end{figure}

As an example for the fits, the light curve for the B 2454826 data (Fig.~\ref{fig:B2LC}) and its rms-magnitude spread (Fig.~\ref{fig:B210}) are shown. A linear fit is superimposed, but it does not
agree with the data. For comparison purposes, a simulated light curve with a log-normal flux distribution is shown in Fig.~\ref{fig:B2sim} (see Section~\ref{sec:testLog} for details of the simulation). It is clear by visual inspection only that the behaviour of the two light curves is very different, a log-normal distribution of fluxes leads to a characteristic behaviour of the peaks as compared to the mean (see Fig.~(4) of \cite{art:5u} and the discussion pertaining to that Figure).

\begin{figure}[!ht]
\includegraphics[scale=0.7]{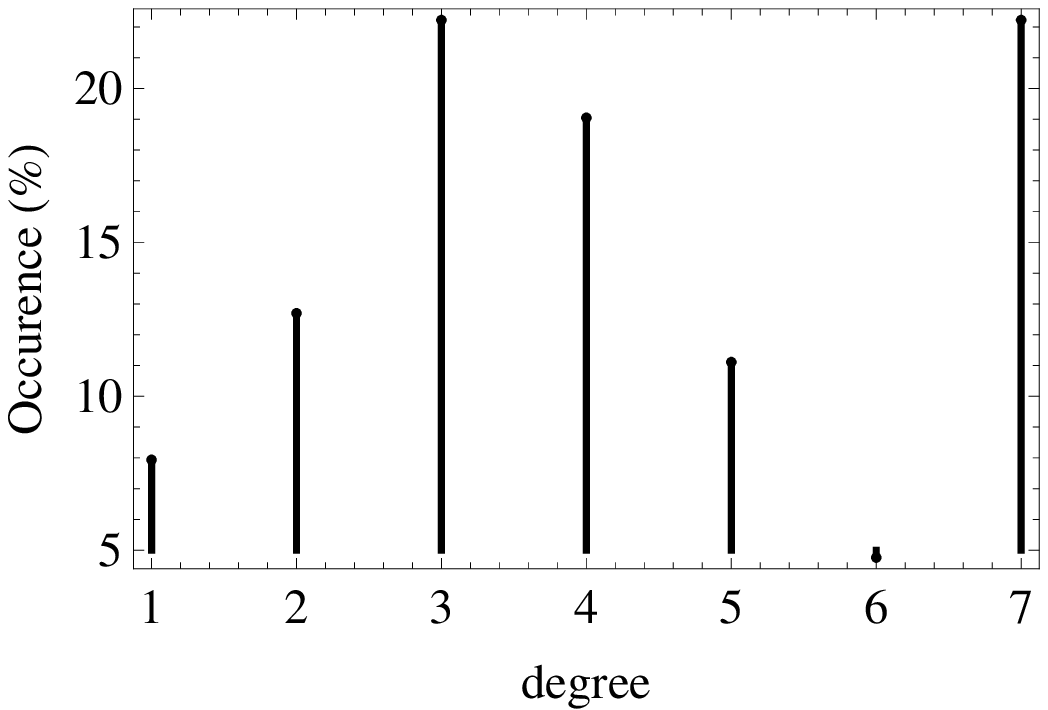}
\caption{Histogram plot of the incidence of polynomial degrees
which fit the data (according to our assumptions and to our data;
the designation "7" stands for the cases were no polynomial fit was possible).} \label{fig:hist}
\end{figure}

A useful way to summarize the main results of the analysis
performed so far
(Tables~\ref{tab:fitsSimultaneous}-\ref{tab:fits3}) is to
visualize the number of data series well fitted by various order
degrees polynomials in a histogram plot (Fig.~\ref{fig:hist}).
Only less than $8\%$ of the data show a linear rms-flux relation with $R^2 > 65\%$.

If the variability timescale $\tau$ is defined as the shortest time between two points, $x_1$ and $x_2$, within a time series such that $\vert x_1 - x_2 \vert \geq 3 \sigma$, the variability timescales for the timeseries investigated in this paper correspond to $\tau \in [1.5, 3.5]\mbox{ hours}$. With a value of the mass of the central object of $M = 10^8 M_\odot$ \cite{art:12f} and with the assumption that the perturbation was produced in the disk, this means that the source of perturbation is placed at a radius $\in [5.49 r_g, 12.81 r_g]$. Correspondingly, we find the following values for natural timescales of the disk \cite{book:10f}: the light crossing time $t_l \in [1.42, 3.33]$ hours, the dynamical timescale $t_{dyn} \in [30.87, 110.02]$ hours and the thermal timescale $t_{th} \left (\alpha _{SS} = 1 \right ) \in [5.91, 21.08]$ years or, more realistically, $t_{th} \left (\alpha _{SS} = 10^{-2} \right ) \in [591.72, 2109.03]$ years. Due to lack of periodicity in IDV, we were not expecting the dynamical timescale to fit the observations and it does not. The standard disk thermal timescale is much longer than the timescales probed in this paper and the standard disk viscous timescale would be even longer. However, the timescales of the standard disk are deduced based on a set of assumptions (see, e.g., \cite{book:13}, Chapter 7). It has been shown that relaxing the no-torque condition on the inner boundary of the disk causes various annuli of the disk to "communicate" on light crossing timescales~\cite{art:20a} and in our case the calculated light-crossing time fits the timescale of the observations. It is thus conceivable that the source of variability is placed in the disk, but not conclusive.

\subsection{Test for log-normal distribution\label{sec:testLog}}

A conclusive diagnosis tool for our purposes would be to test whether or not the optical magnitude is log-normally distributed. Statistical theory states that if the random variable $y = \log {x}$ is Gaussian distributed and if the central limit theorem holds, than the random variable $x$ is log-normally distributed. A prerequisite for the central limit theorem to hold is that the timeseries has a formally infinite number of points. However, it can be shown that the convergence can be realised for a smaller number of points. Even so this condition is restrictive and we cannot perform this type of analysis for all the timeseries described in Tables~\ref{tab:fitsSimultaneous}-\ref{tab:fits3}. 

\begin{table}
\caption{Results for the test of the hypothesis that the timeseries are log-normally distributed, with $\chi ^2 _{ref} = 11.07$, eight bins and five degrees of freedom.} \label{tab:logNorm}
\begin{tabular}{cccc}\hline
   Date & Band & $\chi ^2$ (exp) & $\chi ^2$ (sim) \\ \hline 
   2454825  & R	& 41.1184   & 7.0220 \\ \hline
   2454826  & R  & 24.1224	& 6.8610 \\ \hline
   2454829  & R  & 87.4848	& 3.7912 \\ \hline	
   2454865 - 2454866 & R  & 21.0899	& 1.7466 \\ \hline
   2454871 - 2454872 & R  & 41.1184	& 10.0346 \\ \hline
   2454825  & V	& 51.5006   & 8.2042 \\ \hline
   2454826  & V  & 19.7716	& 11.4255 \\ \hline
   2454826  & B  & 20.2235	& 6.9422 \\ \hline
      
\end{tabular}
\end{table}

\begin{figure}[!h]
\includegraphics[scale=0.4]{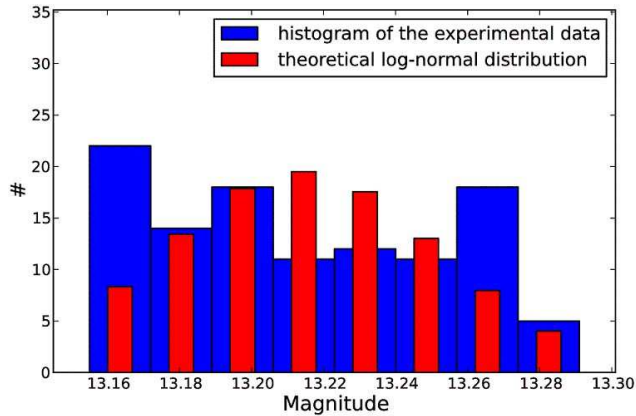}
\caption{histogram of the R 2454825 data with a log normal distribution superimposed. }\label{fig:r6Exp}
\end{figure}

\begin{figure}[!h]
\includegraphics[scale=0.4]{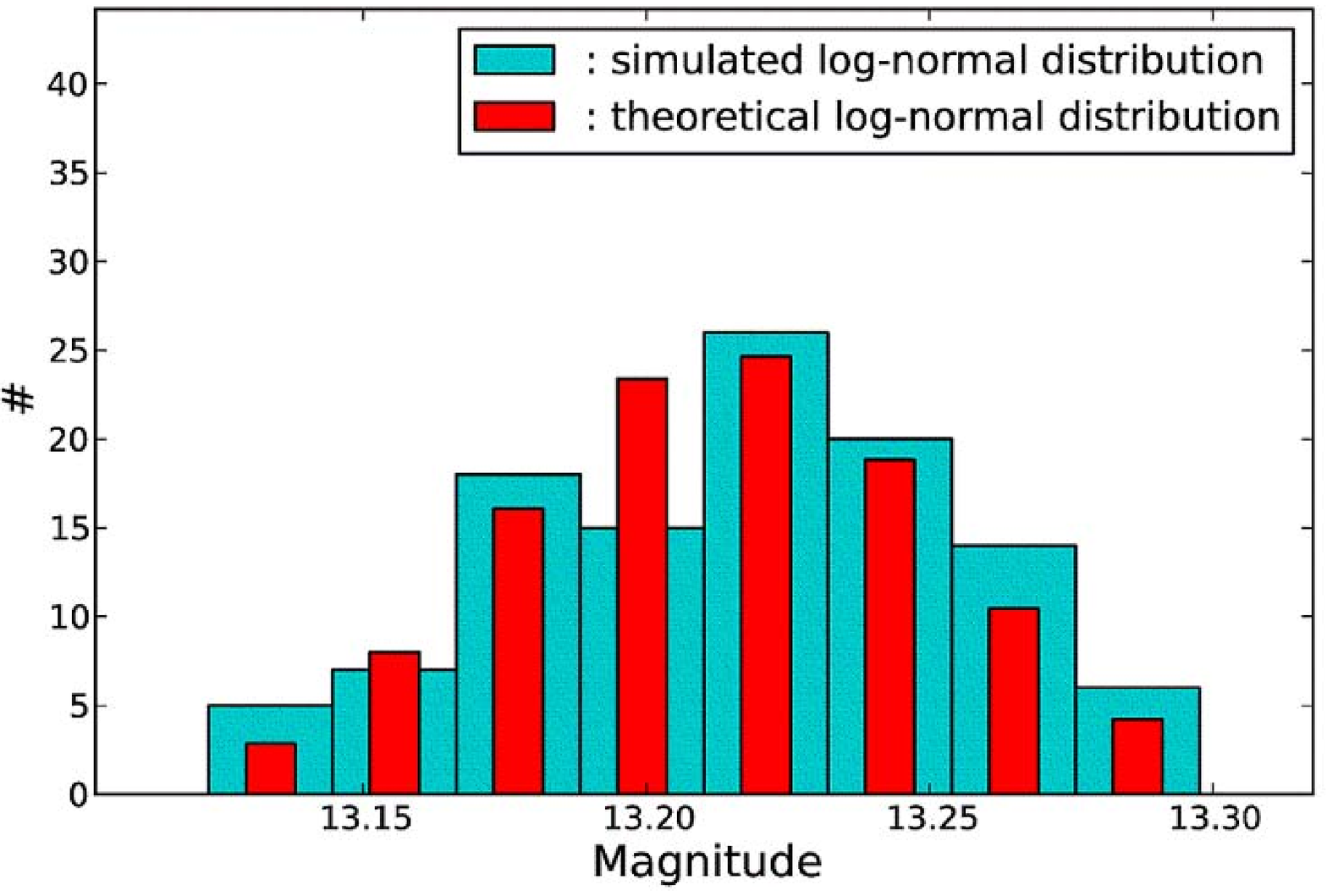}
\caption{simulated log normal distribution for a series with the same mean, variance and number of points as in Fig.~\ref{fig:r6Exp}.}\label{fig:r6The}
\end{figure}

For the purpose of this analysis, we use the chi-square goodness of fit statistics and produce Table~\ref{tab:logNorm}, where the timeseries included have at least $100$ data points. The first and the second column give the identification details of the time series. The third column contains the value of the $\chi ^2$ statistics for the hypothesis that the timeseries is log normally distributed, applied to the experimental timeseries. The forth column contains the same results, but for a simulated timeseries. The constraint that each bin in the histogram has at least five members was obeyed at all times. This forced us to use only eight bins. In turn, for the simulations, out of the maximum eight degrees of freedom of the $\chi ^2$ distribution, three were constrained as follows: the simulated distribution has the same mean, variance and number of points as the observational timeseries. The value of the reference statistics $\chi ^2 _{ref}$ is $11.07$. If $\chi ^2 > \chi ^2 _{ref}$ then the tested hypothesis is rejected with a confidence of $95\%$. None of the time series we could use shows a log normal distribution. However, for the simulated timeseries, $\chi ^2 (sim)$ is consistently smaller than $\chi ^2 _{ref}$. For an illustration of this procedure, the simulated light curve of B 2454826 (Fig.~\ref{fig:B2sim}), the histogram of the R 2454825 data with a log normal distribution superimposed (Fig.~\ref{fig:r6Exp}) and a simulated log normal distribution (Fig.~\ref{fig:r6The}) are shown. 

\section{Conclusions and cautionary notes \label{sect:conclusions}}

The existence of a linear rms-flux relation in 63 time series in
the optical data for the object S5 0716+714 was investigated.
Although some of the data series show an acceptable linear fit,
the general trend is that there is no such linear-rms flux
relation in the optical band (Fig.~\ref{fig:hist}).

We also checked if the distribution of magnitudes is log-normal (Table~\ref{tab:logNorm} and also Figs.~\ref{fig:r6Exp} and \ref{fig:r6The}). The test was done on a restricted number of timeseries (eight), for reasons explained in the body of the paper. None of these timeseries shows a log normal distribution.

The framework initiated by~\cite{art:29l} naturally explains this
relation in the X-Ray band by assuming independent variations in
the accretion rate at different radii and it is, by its
assumptions, a disk model. If one agrees to extend this argument
to say that all disk variability should then exhibit a linear
rms-flux relation, one can say that the optical IDV variability
analysed in this work is not of disk nature.

However, it is obvious that this preliminary result should be taken with the necessary cautionary notes. The rms-magnitude behaviour might prove to be very different for longer timescale variations. It is thus necessary that all groups of timescales be analysed in our proposed framework. Because of the binning procedure, a better statistics would be obtained if the length of each individual timeseries would be a generous multiple of $10$. The probability of IDV detection itself increases with observation time (see, e.g., \cite{art:22r}). Longer timeseries are also needed for the proper testing for a log-normal distribution of the magnitudes/fluxes. In order to confirm or disprove our initial analysis for BL Lac S5 0716+714 and to add a greater degree of generality to the result by including other objects in the analysis, all of these notes must be taken into account in the future.

\subsubsection*{Acknowledgements}

GM gratefully acknowledges the support of Prof. Tiberiu Harko
throughout the completion of this study and Dr. Leung Chun-Sing
for providing some of the data. GM thanks the Institute for
Theoretical Physics, Vienna University of Technology, Austria, for
their hospitality during the time when this work was drafted. The authors are grateful to the anonymous referee for the insightful comments and suggestions that helped to bring major improvements to this work. GM is supported by the project POSDRU/ 107/ 1.5/ S/ 76841. The work of S.B. was supported by a Research Excellence bursary of the BBU.

\bibliographystyle{plain}
\bibliography{AGNrefs}

\end{document}